\documentclass[]{aastex62}
\usepackage{longtable}
\usepackage{multirow}

\newcommand\etal{\it{et al.} \rm }
\accepted{Sep 30, 2019}
\submitjournal{ApJ}

\shorttitle{Spin parity}
\shortauthors{Iye \etal}

\begin{document}

\title{Spin Parity of Spiral Galaxies I -- Corroborative Evidence for Trailing Spirals}

\correspondingauthor{Masanori Iye}
\email{m.iye@nao.ac.jp}

\author{Masanori Iye}
\affil{National Astronomical Observatory of Japan, Osawa 2-21-1, Mitaka, Tokyo 181-8588 Japan}
\affil{National Institutes of Natural Sciences, Hulic Kamiyacho Building 4-3-13 Toranomon, Minato,  Tokyo 105-0001 Japan}

\author{Kenichi Tadaki}
\affil{National Astronomical Observatory of Japan, Osawa 2-21-1, Mitaka, Tokyo 181-8588 Japan}

\author{Hideya Fukumoto}
\affil{The Open University of Japan, 2-11 Wakaba, Mihama-ku, Chiba 261-8586 Japan}



\begin{abstract}



Whether the spiral structure of galaxies is trailing or leading has been a subject of debate. 
We present a new spin parity catalog of 146 spiral galaxies that lists the following three pieces of information: whether the spiral structure observed on the sky is S-wise or Z-wise; which side of the minor axis of the galaxy is darker and redder, based on examination of Pan-STARRS and/or ESO/DSS2 red image archives, and which side of the major axis of the galaxy is approaching to us based on the published literature.
This paper confirms that all of the spiral galaxies in the catalog show a consistent relationship among these three parameters, without any confirmed counterexamples, that supports the generally accepted interpretation that all the spiral galaxies are trailing and that the darker/redder side of the galactic disk is closer to us.  Although the results of this paper may not be surprising, they provide a rationale for analyzing the S/Z winding distribution of spiral galaxies, using the large and uniform image data bases available now and in the near future, to study the spin vorticity distribution of galaxies in order to constrain the formation scenarios of galaxies and the large-scale structure of the Universe.

\end{abstract}




\section{Introduction} 
\subsection{Observational Studies}

The sense of winding of spiral arms in disk galaxies with respect to their rotation was a classic subject of debate in observational and theoretical studies of spiral galaxies until the 1960s.

\citet{Slipher1917} and \citet{Curtis1918} noted the asymmetric distribution of dark lanes in many of spiral galaxies and argued that the more obscured, redder side of the minor axis is likely nearer to us. \citet{Slipher1917} even remarked that all spiral nebulae rotate in the same direction with reference to the spiral arms, as a spring turns in winding up. 
This type of spiral is called a trailing spiral.  As the inner part of a galaxy revolves at a larger angular speed than the outer part, the trailing pattern will wind up in the course of time if the spiral arms are material arms. 
\citet{Hubble1943} studied 15 spiral galaxies and concluded that all of them are trailing spirals, assuming that the darker side of galaxies is closer to us.  

In contrast to the interpretation of trailing spirals, \citet{Lindblad1940a} and \cite{Lindblad1940b} developed a dynamical theory of the spiral structure of stellar systems. He argued that the direction of rotation of the system agrees with the direction in which the spiral arms wind around the center when followed outward from the central mass, suggesting that the spirals are unwinding.  This type of spiral is called a ``leading'' spiral. 
\citet{Lindblad1946} further noted that the bright side of the tilted disk of spirals is the near side.
 They argued that the obscuring dust has a larger vertical extension than that of the luminous matter, and hence produces a ``limb-darkening'' effect on the far side of the tilted disk, because of the longer path length for obscuration.  This interpretation implies that the spirals are ``opening up''. As the thickness of the distribution of the obscuring dust is now known to be much  smaller than that of the luminous stellar system \citet{Hacke1982}, this historic argument is no longer valid.   

Even if admitting the vertically thin distribution of obscuring matter, two different interpretations were still possible.  Figure 1, reproduced from a review article by \citet{de Vaucouleurs1959}, illustrates the two interpretations.  The upper panel shows usual interpretation that the dust layer is more or less uniformly spread in the equatorial plan of the disk and the obscured side is the near side.  The lower panel shows Lindbland's interpretation that the dust lanes localized in the inner edge of the bright arms and the obscured side is the far side.

Nevertheless, it is now generally accepted that the dust-lane-dominant side of the minor axis is the near side.  Here dark lanes are more conspicuous on the side nearer to us because the interstellar dust lanes are silhouetted against the background light of the stellar bulge, as \citet{de Vaucouleurs1959} suggested.  In the absence of direct distance measurement of the near and far sides of tilted galactic disks, however, this interpretation remains incompletely verified by other independent methods, e.g. confirming the differential interstellar reddening of globular clusters and/or the detection of differences between forward and backward scattering of light from the stellar bulge by the interstellar dust using observations in polarized light and/or observations of  spectral differences.

To  identify whether a spiral arm of a galaxy is trailing or leading in a galaxy, one needs to know the projected spiral winding sense (S-wise or Z-wise) on the sky, the near side of the tilted galactic disk, and the approaching side of the major axis of the galactic disk.

Determining whether a spiral pattern exhibits S or Z winding is difficult for highly inclined galaxies, whereas for nearly face-on galaxies, it is hard to identify the near side of the disk from the asymmetric extinction.  It must be emphasized that to date there is yet no spiral galaxy for which all three pieces of relevant information are indisputably available.   

\begin{figure}[h]
\centering
\includegraphics[scale=0.5]{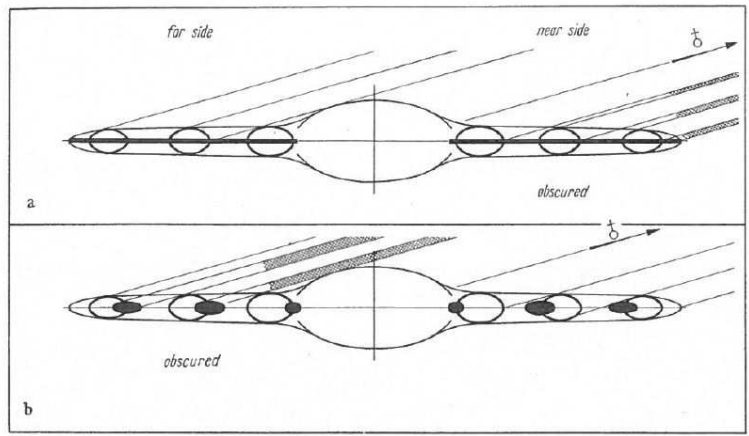}
\caption{Two geometrical configurations for the distribution of obscuring dust layer in a tilted disk of a spiral galaxy.  (a) Usual interpretation; dust-lane-dominant side is the near side. (b) Lindblad's interpretation; dust-lane-dominant side is the far side (reproduced from \citet{de Vaucouleurs1959}.}
\label{DeVau1959}
\end{figure}
\subsection{Theoretical Studies}

\citet{Hunter1963, Hunter1965}, \citet{Iye1978}, and \citet{Kalnajs1972} showed analytically that no spiral modes arise in uniformly rotating gaseous or stellar disks.  This is called the ``anti-spiral'' theorem. Differential rotation is the key to spiral modes.  One can formulate the dynamical stability of a self-gravitating disk as an eigenvalue problem of the dynamical response  matrix of  a disk galaxy. The eigenvalues of such a matrix with real components appear only in real values or in complex conjugate pairs.  Consequently, if there is a growing trailing spiral mode, there is a corresponding decaying leading spiral mode.  The actual spiral patterns observed in real galaxies can be regarded as a manifestation of superposition of these various modes of oscillation, at least in the linear regime.

\citet{Lin1964} developed the density wave theory to explain the sustained existence of spiral structure as a manifestation of density waves rather than  a material structure to avoid the so-called ``winding dilemma''. They formulated the governing equations of local linear perturbations from circularly rotating equilibrium state of a disk galaxy and derived the dispersion relation for waves. \citet{Toomre1969} studied the propagation of such density waves across the galactic disk and found that leading arms are unwinding and turn into trailing arms which are wound up more tightly. He called this excitation mechanism of spiral ``swing amplification.''

\citet{Hunter1969} studied short-wavelength oscillation of cold, thin, gaseous disk models of galaxies using a WKB approximation and concluded that leading spirals are gravitationally unstable and growing.  Actual galaxies are not completely cold, and the dispersion equation of linear perturbation shows that short waves are easily stabilized by introducing a modest pressure or velocity dispersion of stars.  

\citet{Hunter1965}, \citet{Bardeen1975}, \citet{Iye1978}, and \citet{Aoki1979}  formulated the global dynamical stability of self-gravitating disk galaxies as an eigenvalue problem by expanding the linear perturbation in terms of a bi-orthogonal set of solutions of the Poisson equation on simple disk models.  The global modal approach to studying the gravitational stability of cool or warm gaseous disks shows that trailing spiral modes, in addition to bar-like modes, are generally growing. 

\citet{Kalnajs1977} developed a rigorous formulation to study the modal stability of stellar disk systems. \citet{Evans1998} extended the method to study the global spiral modes of power-law disks.  \citet{Jalali2005} made an extensive study of the global spiral modes of the Kuzmin disk (\citet{Kuzmin1956}), Miyamoto disk (\citet{Miyamoto1971}) and  isochrone disk.  They demonstrated that trailing spiral modes are growing and that they transfer angular momentum from the interior outward. 

In addition to these analytical studies, N-body numerical simulations on the dynamics of disk galaxies have shown from an early stage the development of trailing spirals (e.g., \citet{Hohl1970}).  \citet{Sellwood1986} and many other works over decades of computer simulation studies have reached the commonly accepted view that the spiral structure in galaxies is trailing.   In summary, there is a widely adopted theoretical expectation that the spiral structure observed in galaxies is trailing. 

The following sections present a new view of this issue using a spin parity catalog of spiral galaxies by examining the current databases of images with consistent orientations and accumulated published results of kinematic observations. 

In Section 2, we describe the image data we used to evaluate the spiral winding sense and the dust-lane-dominant side of each spiral galaxy.  We also describe the major surveys from which we retrieved the information to identify the approaching side.  Section 3 describes the sample selection and Tables 2-5 which are the outcome of the present study.  Table 2 shows 146 spirals for which we were able to confirm the three pieces of relevant information.  It clearly shows that all the spiral galaxies studied are trailing spirals, providing the main result of this paper.  Tables 3-5 show additional catalogs for which only two of the three pieces of information were observationally confirmed.  Our study makes it possible to infer the third piece of information by assuming that all the spiral galaxies in the Universe are trailing spirals.  Section 4 discusses some individual galaxies and our plan to use the present findings to study the vorticity distribution of galaxies in the Universe without recourse to spectroscopy.


\section{Spin Parity of Spiral Galaxies}

\subsection{Projected Sense of Winding of Spiral Arms}  \label{subsec:winding}

For most spiral galaxies with morphological type classifications of Sb to Sd, the projected sense of winding of their spiral arms can be evaluated in a fairly straightforward manner.  It must be noted, however, that some of the images shown on web sites, in the literature and in books are often inverted.  For instance, the image of NGC5055 in the classical ``Atlas of Galaxies Useful for Measuring the Cosmological Distance Scale'' by \citet{Sandage1988} is inverted, whereas the images of the same galaxy in \citet{Sandage1994} and \citet{Sandage1961} show the galaxy correctly as seen in the sky.  We evaluated whether the spiral winding direction on the sky is S-wise or Z-wise (reverse S-wise) from the consistently oriented images taken from PanSTARRS-1 data or ESO/DSS red data to avoid the potential confusion arising from such image inversion.  

We adopted mainly the \it{y/i/g} \rm color composite images from PanSTARRS-1 database. For some galaxies where the \it{y/i/g} \rm color composite image has some defects, we adopted either of \it{i-} \rm or \it{g-} \rm band images that shows the dust-lane-dominant side and spiral features most clearly.  When PanSTARRS-1 images were not available, we used ESO/DSS2 red images.   We did not apply quantitative limits to their inclination angle of galactic disks to select and evaluate the spiral winding direction or dust-lane-dominant side. Figures 2 and 3 show typical examples of spirals with S-wise and Z-wise winding.

\begin{figure}[h]
\begin{tabular}{c|c}
\begin{minipage}{0.45\hsize}
\centering
\includegraphics[scale=0.2]{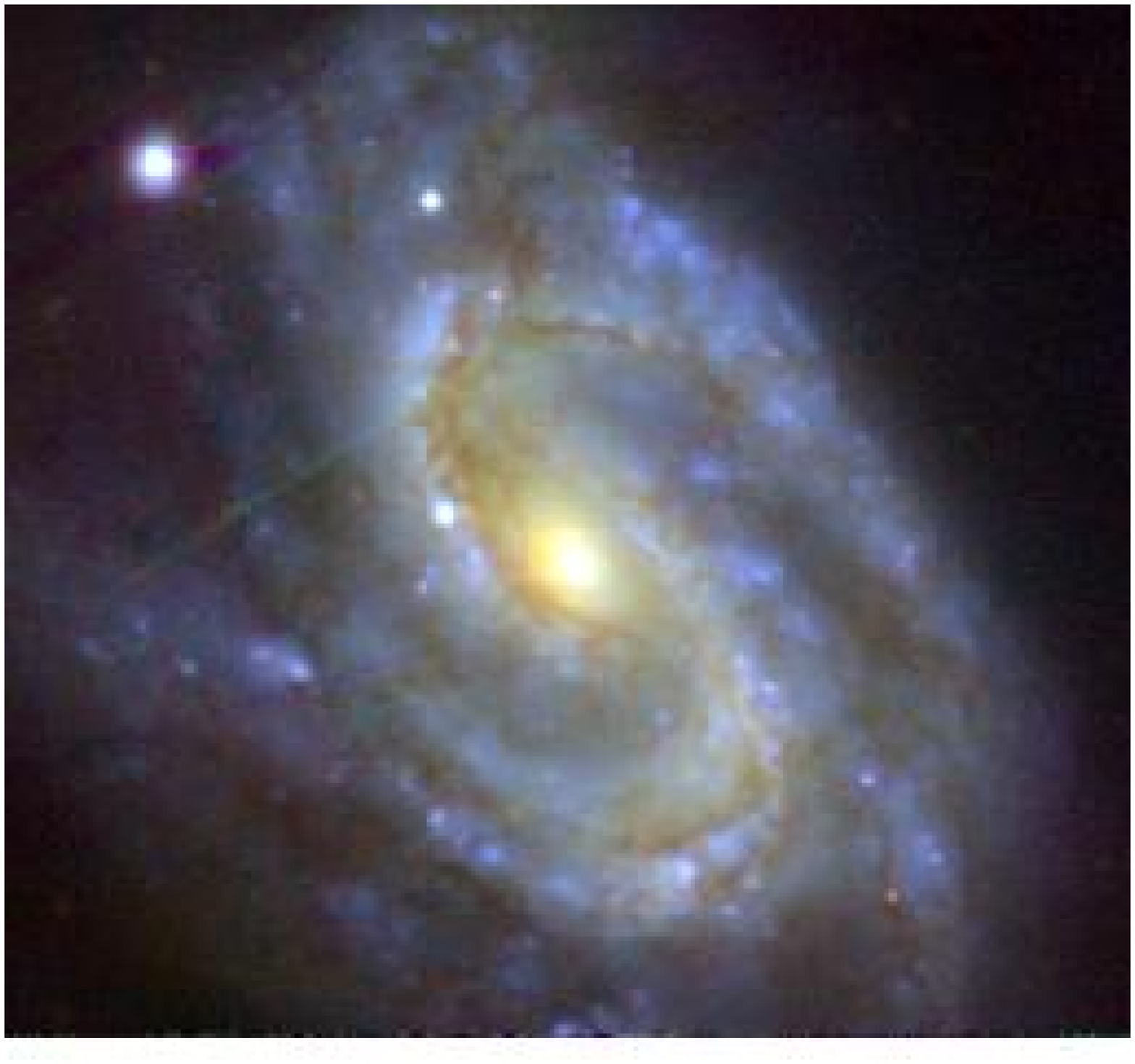}
\caption{S-wise winding spiral NGC157}
\end{minipage}
\begin{minipage}{0.45\hsize}
\centering
\includegraphics[scale=0.2]{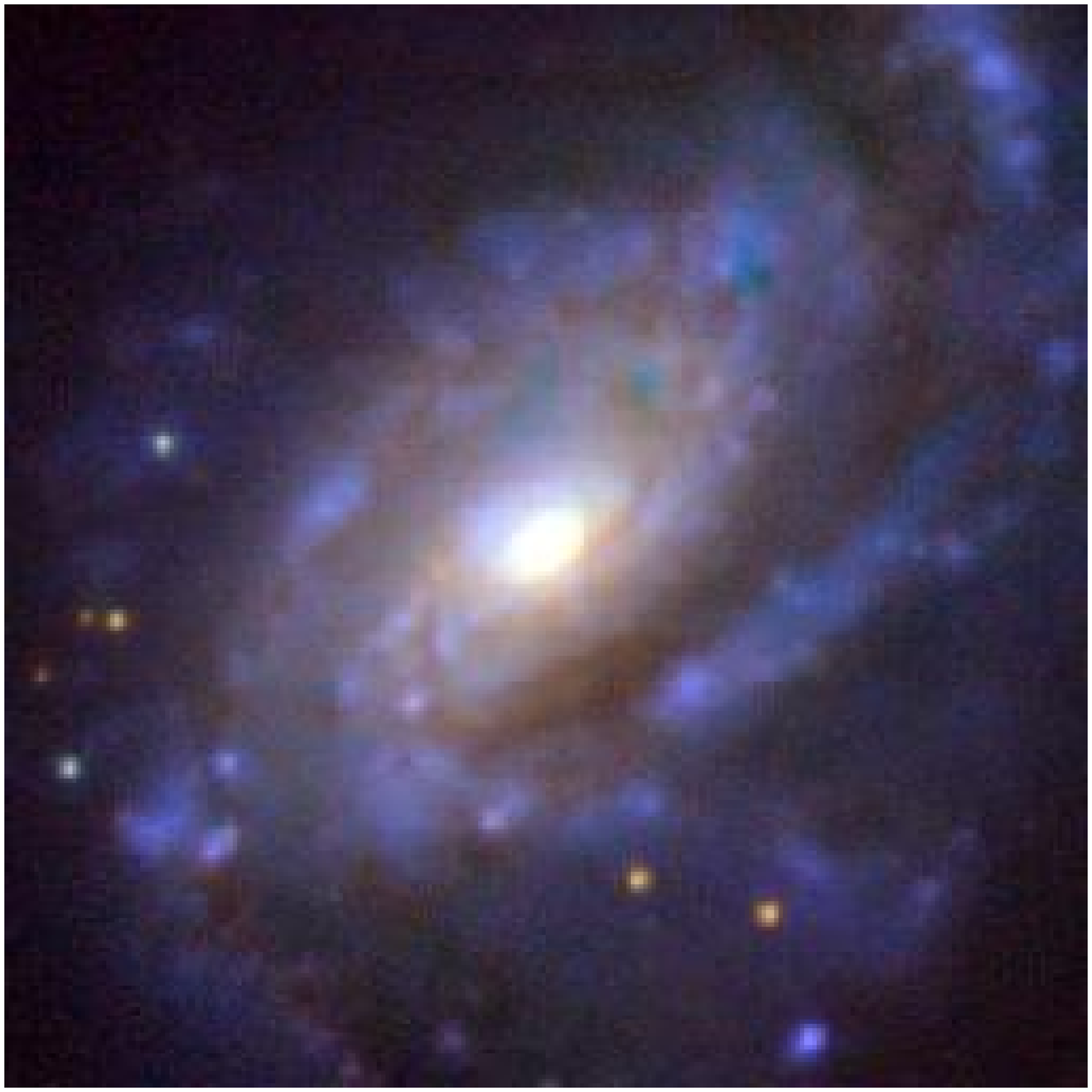}
\caption{Z-wise winding spiral NGC877}
\end{minipage}
\end{tabular}
\end{figure}

For NGC224, NGC 1772 and NGC 4622, we employed images from other sources, as noted in section 4, to evaluate the spiral winding direction.  Spiral features with opposite winding directions reportedly coexist in a few galaxies, which are noted in section 4.

\subsection{Dust-Lane-Dominant Side on the Minor Axis} \label{subsec:dark side}

Interstellar dust lanes condensed mainly on the disk plane in front of the bulge are silhouetted against the background light of the stellar bulge, whereas the dust lanes behind the bulge are not clearly visible.  Therefore, the dust-lane-dominant side of the tilted disk is generally considered to be nearer to us, as suggested by \citet{Slipher1917}.  As the vertical thickness of the dust lanes is smaller than that of the stellar disks, this effect is still noticeable for some spiral galaxies with smaller bulges.

This notion is generally consistent with models that explain in detail the differential reddening and dust obscuration between the near and far sides of a galaxy, for example, \citet{Elmegreen1999} or \citet{Byun1994}.

However, this interpretation has not been fully verified by independent observational methods.   Measurement of differential interstellar reddening of globular cluster systems can be used to independently assess the correctness of this assumption.  Globular clusters behind the galactic disk should be statistically redder than those in front of the galactic disk as a result of the reddening effect of interstellar dust in the galactic disk. To date, this has been clearly demonstrated only for M31 (\cite{Iye1985}), but no similar result has been reported for other galaxies. For M31, \citet{Iye1985} showed that the northwest side, where the dust lanes are more conspicuous than on the southeast side of the disk, is actually the near side.

Differential reddening in different parts of the disk has also been observed in other local galaxies and used to deduce the geometry and near side of the disk, for example, in M33 (\citet{Cioni2008}) and in the LMC (\citet{van der Marel2001}.

In this paper, where possible, we visually identified the dust-lane-dominant side of the minor axis of the galactic disk  Figures 4 and 5 show two examples.

\begin{figure}[h]
\begin{tabular}{cc}
\begin{minipage}{0.45\hsize}
\centering
\includegraphics[scale=0.2]{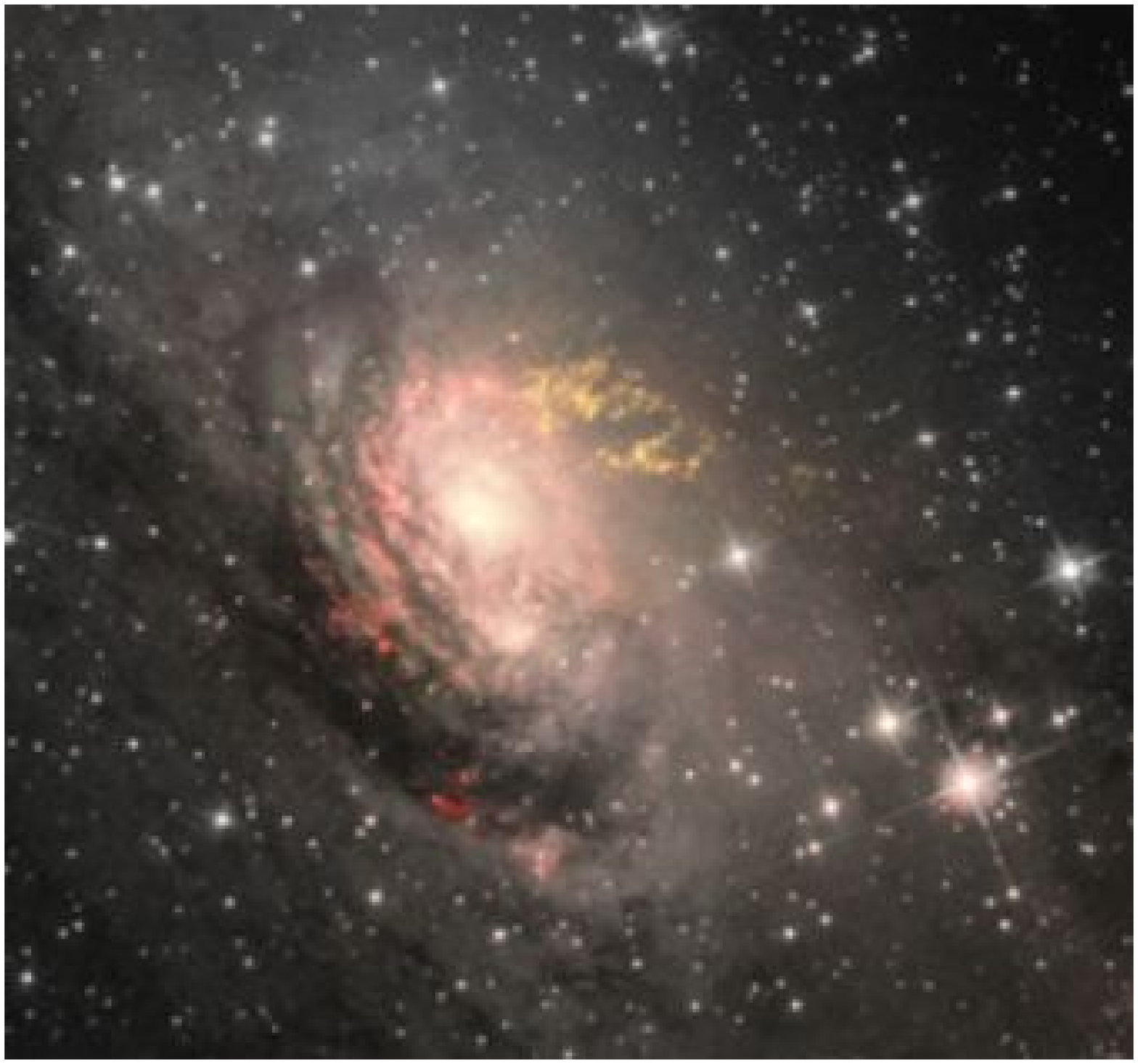}
\caption{Circinus galaxy clearly shows more conspicuous dust lanes on its southeast (lower left) side.}
\end{minipage}
\begin{minipage}{0.45\hsize}
\centering
\includegraphics[scale=0.2]{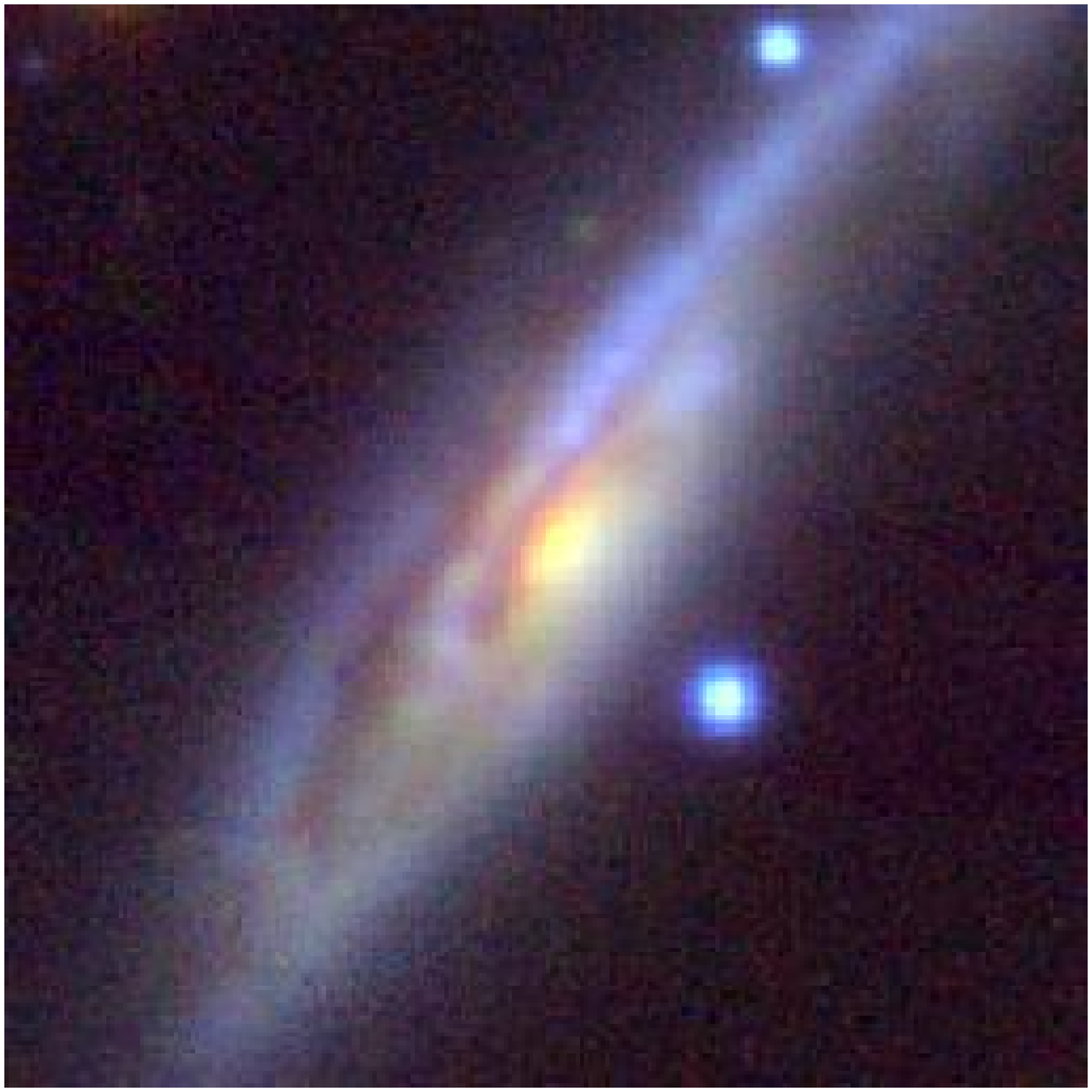}
\caption{Northeast side of IC2101 shows dust lanes more clearly than on northeast (upper left) side.}
\end{minipage}
\end{tabular}
\end{figure}

\subsection{Approaching Side of the Major Axis} \label{subsec:approaching}

Published data from several useful systematic observational studies such as the Gassendi H-Alpha survey of Spirals (GHASP survey; \citet{Epinat2010}, \citet{Korsaga2018}, and \citet{Moretti2018}), PPak IFU survey (\citet{Martinsson2013}), WIYN Telescope H-$\alpha$ \rm IFU survey (\citet{Andersen2013}) Westerbork Synthesis Radio Telescope HI survey (\citet{den Heijer2015}), Calar Alto Legacy Integral Field Area, CALIFA survey (\citet{Garcia-Lorenzo2015}, \citet{Holmes2015}, and \citet{Falcon-Barroso2017}), and VLT/SINFONI integral field spectroscopy (\citet{Mazzalay2014}) were compiled in the present study.  Other published literature was searched using ADS with keywords of ``rotation curve'' or ``velocity field'' coupled with ``spiral galaxies'' or the explicit names of individual galaxies to confirm the approaching side on the major axis of the galactic disks.

\begin{figure}[h]
\begin{tabular}{cc}
\begin{minipage}{0.45\hsize}
\centering
\includegraphics[scale=0.2]{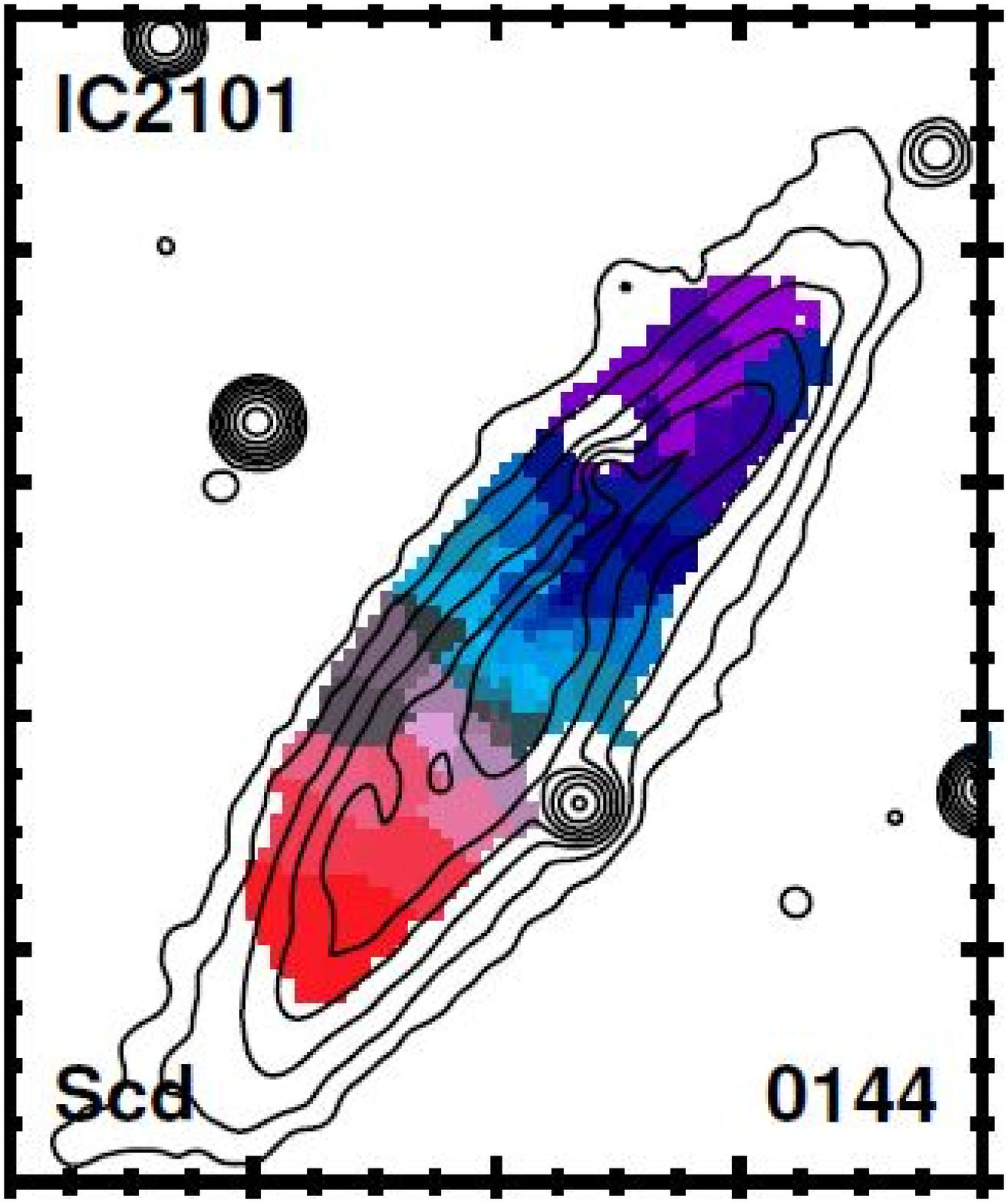}
\caption{CALIFA iso-velocity map of IC2101 shows that the northwest (upper right) side is approaching (\citet{Falcon-Barroso2017}).}
\end{minipage}
\begin{minipage}{0.45\hsize}
\centering
\includegraphics[scale=0.4]{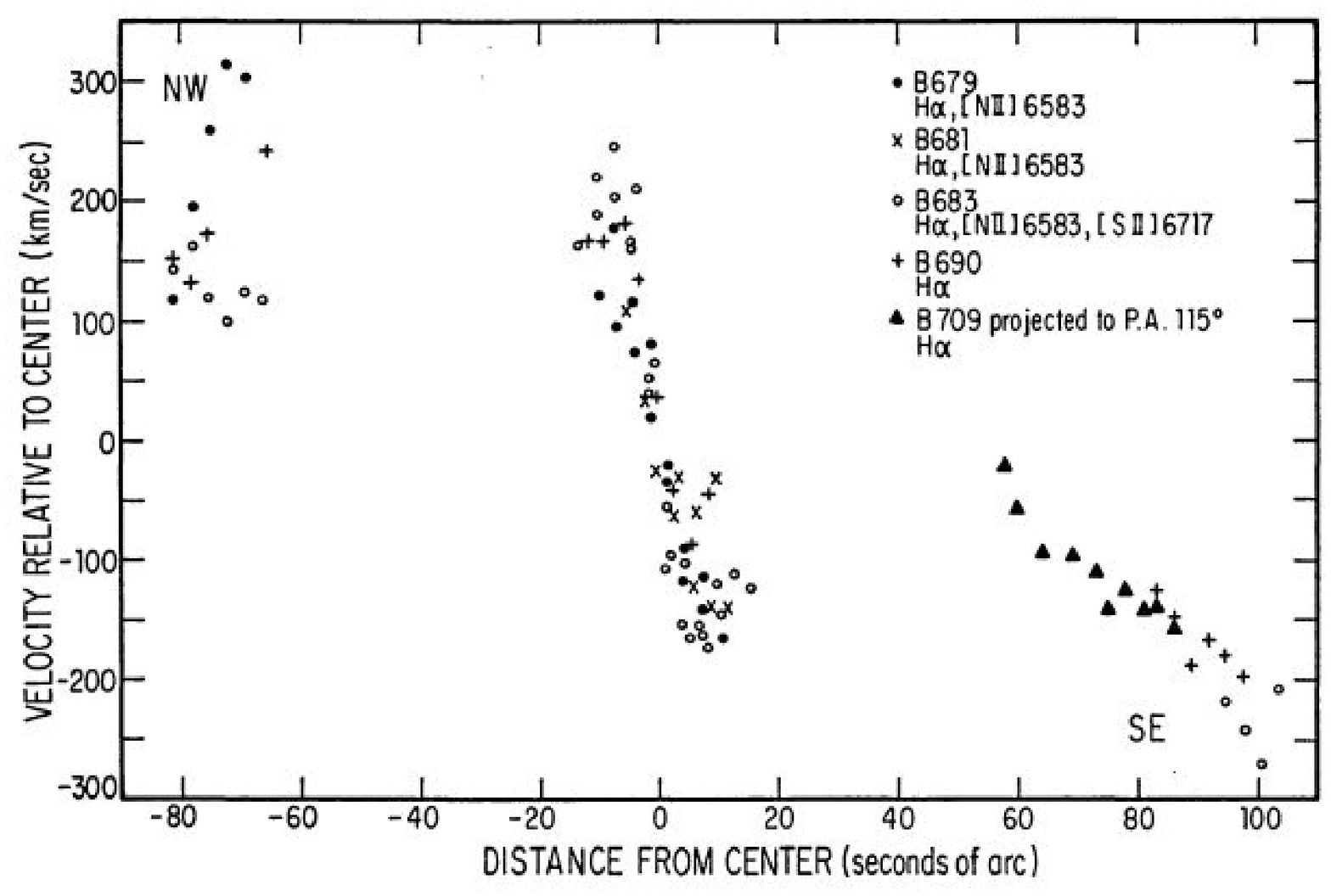}
\caption{Kinematic study NGC613 shows that  the southeast side is approaching (\citet{Burbidge1964})}
\end{minipage}
\end{tabular}
\end{figure}

We made every effort to confirm that the maps used to establish the approaching and receding sides have the same orientation as the optical images.
Only reports that explicitly describes the approaching side of the major axis were used. Note that some papers use folded rotation curves as a function of the distance from the center or rotation curves, showing approaching and receding velocities, but did not provide sufficient information on the direction of the approaching side.  These papers were not used in the present study to determine the approaching side.

Figures 6 and 7 show examples of a velocity field map and a rotation curve, from which we identified the approaching side of the disk galaxy, respectively. The papers on galaxy kinematics used to infer the approaching side of the galaxies are listed in column (6) of Table 2 and in column (5) of Table 3.

\section{Spin Catalog} \label{sec:catalog}

We first compiled a master list of spiral galaxies, from the published literature, with available kinematical data to identify the approaching side.  The literature survey was performed on ADS by searching on various relevant keywords.  We then looked at images of these galaxies in the Pan-STARRS and/or ESO/DSS2 image archives to see if we can determine the spiral winding sense and dust-lane-dominant side.  During this process we added spiral galaxies for which relevant information on spiral winding sense and/or dust-lane-dominant side can be retrieved from their images even though the kinematical data are absent.   Image inspection processes were done independently by two of the authors, MI and HF.  Only those galaxies for which both authors obtained the same spiral winding direction and dust-lane-dominant side were chosen for study. 

The master list eventually contained 845 spiral galaxies, of which S/Z identification was made for 784 spirals, approaching side identification for 520 spirals, and dust-lane-dominant side identification for 245 spirals, respectively. Table 1 shows the statistics of the master list samples.   The numbers of galaxies for which only one piece of information was available are also shown; 252 spirals with only S/Z identification, 8 spirals with only dark side identification and 27 spirals with only approaching side identification.  The catalogs for these spirals with insufficient information are not included in this paper. \\
\begin{table}[h]
\begin{center}
\caption{Number of spiral galaxies examined} 
The check marks indicate that the judgment result is available for (1) S/Z spiral winding, 
(2) dust-lane-dominant side, denoted here dark side, and (3) the approaching side.\\

\begin{tabular}{|c|c|c|c|} \hline
(1)&(2)&(3)&(4) \\
S/Z Winding & Dark Side & Approaching Side & Number of Spirals \\ \hline
$\surd$ & $\surd$ & $\surd$ & 146 \\
$\surd$ &   -   & $\surd$ &  321 \\
$\surd$ & $\surd$ &   -   & 63 \\
   -   & $\surd$ & $\surd$ & 25 \\
$\surd$ &   -    &    -  & 252 \\
   -    & $\surd$ &  -    &  8 \\
   -   &   -    & $\surd$ & 27 \\ \hline
782 & 242 & 519 & 842 \\ \hline
\end{tabular}
\end{center}
\end{table}

Table 2 lists the final sample of 146 spiral galaxies for which the winding direction (S/Z), dust-lane-dominant side, and approaching side are clearly confirmed.
Column (1) lists the ID of each galaxy. Column (2) gives the spiral winding sense, S-wise (S) or anti-S-wise (Z), as projected on the sky.  Column (3) is the sky cardinal direction of the dust-lane-dominant side along the minor axis.  Column (4) shows the sky cardinal direction of the approaching side of the major axis.  Column (5) shows whether the spiral structure is trailing (T) or leading (L) based on columns (2)-(4) where we assume that the dust-lane-dominant side is on the near side of the galactic disk.  Column (6) is the reference we used to confirm the approaching side.  Columns (7) - (10) show postage-stamp images of four galaxies that cover the main spiral features of the galaxy; most of them came from the \it{y/i/g} \rm color composite images of PanSTARRS-1 Image Cutout Server (https://ps1images.stsci.edu/cgi-bin/ps1cutouts).     If PanSTARRS-1 images of a galaxy were not available, we used the ESO/DSS-2 red images (http://archive.eso.org/dss/dss).  

Two of the authors (MI and HF) visually inspected the images to classify the galaxies as S or Z and determined the direction of the dust-lane-dominant side.  In visually judging S/Z, we relied more on the main outer features than on local features. Only when MI and HF confidently obtained the same results was a galaxy selected for study.   
The approaching side was identified using iso-velocity contours or the velocity curves in the references in column (6) of Table 2 and column (5) of Table 3.
Ultimately we needed only the three relevant pieces of information, so we did not quantify the position angle.

Note that the distribution of eight cardinal directions assigned here is not uniform.  Where possible, we assigned intercardinal directions (NW, NE, SE and SW) rather than cardinal directions (N, E, S, and W). We did this to ensure self-consistency of the orthogonality of the directions of the dust-lane-dominant side and the approaching side, especially for spirals with their small inclination angles.  
HH
Note also that the decision on whether the spiral is the trailing or leading type [column (5)] was made only after all the information in columns (2)–(4) was determined so that the information in columns (2)–(4) was not affected by the information in column (5).

Remarkably, all 146 galaxies have trailing spiral arms if the dust-lane-dominant side is closer to us.  The decisions for only two of the 146 spiral galaxies, NGC 722 and NGC 4622, are somewhat questionable, as will be discussed in section 4.  It has been widely assumed that all the spiral galaxies are trailing, but this paper provides an up-to-date observational confirmation of this generally accepted idea.

Table 3 lists 321 spiral galaxies for which the spiral winding direction [column (2), uppercase letters] and the approaching side [column (4), uppercase letters] are taken from the literature [column (5)]. Individual postage-stamp images are shown in columns (7)–(10), as in Table 1.  Column (3) shows, in lowercase letters, the inferred near and dark sides of the minor axis of each galaxy, assuming that all the spirals are trailing.

Table 4 lists the spiral winding direction column (2), uppercase letters] and dark side on the minor axis [column (3), uppercase letters] of 63 spiral galaxies.  The inferred approaching side of these spiral galaxies, assuming that all the spirals are trailing, is given in lowercase letters in column (4).  

Table 5 lists 25 galaxies for which the darker side [column (3), uppercase letters] and the approaching side [column (4), uppercase letters] are observationally confirmed.  The inferred sense of spiral winding, assuming that all the spirals are trailing, is given in lowercase letters in column (2). 

Table 6 lists other names of some of the sample galaxies.

\section{Discussion} \label{sec:discussion}

\subsection{Notes on Individual Galaxies} \label{subsec: notes}

\noindent IC 2101 \hspace{5mm} \\
IC 2101 is clearly redder on the northeast side and shows S-type spiral structure. \\

\noindent NGC224 \hspace{5mm} \\
NGC224 (M31) is known to rotate with its southwest side approaching us (\citet{Peace1918}, \citet{Babcock1938}).  A publicly available color image is adopted to fit in Table 2.  The northwest side of the disk, where the dust lanes are more conspicuous than on the southeast side, was confirmed to be the near side by a differential interstellar reddening study of its globular cluster system (\citet{Iye1985}).  This galaxy remains the only galaxy for which there is firm independent proof of the generally adopted idea that the dust-lane-dominant dark side is the near side.  Whether M31 has S-wise or Z-wise spiral winding as projected on the sky has been somewhat controversial because of the rather high inclination of the galactic disk plane.  

\citet{Simien1978} examined the distribution of 981 HII regions in M31 and concluded that a one-armed leading spiral fits the data better than a classical two-armed trailing spiral.  However, we used the face-on image produced by Efremov (2009) to determine that M31 has S-type two-armed spiral structure rather than a Z-type spiral. This concludes that M31 is a trailing spiral. \\

\noindent NGC772 \hspace{5mm}  \\
The figures in both \citet{Pignatelli2001} and \citet{Rhee1996} show that the northwest side is approaching.  This makes NGC772 a unique galaxy with leading spiral arms.   Heraudeau and \citet{Simien1998}, however, indicated that the southeast sides is approaching according to their convention of showing the rotation curve for PA=130$^\circ$, \rm and this places NGC772 in the category of trailing spirals.  For NGC772, additional observational confirmation of which side of the major axis is approaching us is necessary to conclude whether the spiral is leading or trailing.\\

\noindent NGC1097 \hspace{5mm} \\
Note that the identification of the near side based on the extinction $A_v$ map obtained from the HST image ratio F814W/F438W of NGC1097 (\citet{Prieto2019}) is consistent with that from the dust lanes.  This confirms that the central spiral of NGC1097 is trailing.  \\

\noindent NGC 2742 \hspace{5mm} \\
NGC 2742 is a good example of a galaxy where the dust lanes are more conspicuous in the color composite image than in monochromatic images.  Note the presence of dust lanes on the southern perimeter of the central bulge. \\

\noindent NGC2775 \hspace{5mm} \\
We determined the Z-wise winding sense of NGC 2775 from high-resolution HST images.  \\

\noindent NGC3124 \hspace{5mm} \\
This galaxy has well-defined S-wise main spiral arms. \citet{Treuthardt2014}, however, pointed out that a Z-wise spiral feature is present at the end of the central bar. \\






\noindent NGC 3160 \hspace{5mm} \\
NGC 3160 is almost edge-on, but we can tell that the southwest side of the disk is nearer to us, as we can see that the dust lane is located slightly southwest of the nucleus of this galaxy.  The Z-wise spiral winding sense for this galaxy is inferred from the extended outer part of the disk.  \\

\noindent NGC4622 \hspace{5mm}  \\
NGC4622 is well-known for its unique spiral structure winding in opposite directions.  The two outer main spiral arms wind S-wise on the sky, but an additional Z-wise spiral arm is visible in the northeastern inner region.  The south side of this galaxy is approaching (\citet{Buta2003}).  
\citet{Thomasson1989} and \citet{Byrd1989} made an N-body simulation of a retrograde encounter by a companion galaxy to reproduce the formation of a leading arm as seen in NGC4622.  \citet{Buta1992} interpreted the inner arm of NGC4622 as a leading arm produced by such an encounter.  \citet{Buta2003} later argued, however, that the east side is the near side on the basis of dust features visible in HST images.  This indicates that NGC4622 is an exceptional galaxy with two main leading spiral arms.  We think, however, the conclusion that the east side of NGC4622 is the near side based on dust lanes is not obvious for this nearly face-on spiral.  We note that dust features are also present on the west side, although they are much fainter than those on the east side.  We prefer, therefore, the interpretation that the west side of NGC4622 is the near side, which allows a more reasonable understanding that the two main spirals are trailing. \\

\noindent NGC5055 \hspace{5mm}  \\
The image of NGC 5055 in the classical atlas of \citet{Sandage1988} is inverted, whereas the images of the same galaxy in \citet{Sandage1994} and \citet{Sandage1961} show it correctly as seen in the sky. \\

\noindent NGC 6015  \hspace{5mm}  \\
NGC 6015 is another good example of a galaxy where the dust lanes are more conspicuous in the color composite image than in monochromatic images.  Note the prevalence of dust lanes on the northwest side of the disk.  \\ 

\noindent UGC 10972  \hspace{5mm}  \\
UGC 10972 is redder on its northwest side, indicating that the northwest side is the near side.  Its S-wise spiral winding is inferred from the outermost arms seen on both ends.  \\

We did not apply quantitative limits on their inclination angle to select and evaluate the spiral winding direction and dust-lane-dominant side.

\subsection{Spiral Pattern as a Tracer of the Spin Vector of Galaxies} \label{subsec: tracer}

This survey of the available data for evaluating the spiral winding direction of disk galaxies shows very clearly that the spiral structure of all the galaxies is trailing if the dust-lane-dominant side of the minor axis of disk galaxies is closer to us.  This study does not find any indisputably confirmed counterexample of leading spirals.  Only two galaxies in our sample, NGC772 and NGC4622, remain somewhat questionable. The approaching side of NGC722 must be confirmed as conflicting results have been reported.  The near side of the nearly face-on spiral NGC4622 is still open to debate.  

The highly reasonable interpretation that the spiral structure seen in disk galaxies is essentially trailing provides us a basis to identify the sign of the line-of-sight component of the spin vector of a spiral galaxy just by identifying whether it is an S-wise spiral or a Z-wise spiral from images.  We can see whether the spin vector of a spiral galaxy is pointing toward us or away from us without obtaining its spectroscopic observational data.
Although it is unlikely, the opposite identification applies if all the spirals are leading instead of trailing. We can still use S/Z identification of spiral galaxies to study if there is any anisotropy in the spin vector distribution of sample galaxies.

The distribution of the sign of the line-of-sight component of the spin vector of spiral galaxies provides a tool to investigate the origin of the vorticity distribution in aggregations of galaxies, including the Local Group, groups of galaxies, clusters of galaxies, and the Local Super Cluster of galaxies.  \citet{Thompson1973}, \citet{Borchkhadze1976}, \citet{Yamagata1981} and \citet{MacGillivray1985} made early attempts to study the S/Z distribution of galaxies to find any bias in the spin parity distribution of galaxies in the northern sky. \citet{Iye1991} studied the S/Z distribution of 8287 spiral galaxies in the southern sky using the ESO/Uppsala Survey of the ESO(B) Atlas.  \citet{Borchkhadze1976} and \citet{Yamagata1981} identified more Sbc-Sc spirals in the northern hemisphere as S-spirals than as Z-spirals, whereas \citet{Iye1991} reported  that Z-spirals were slightly dominant over S-spirals in the southern hemisphere.  \citet{Sugai1995} searched for a spatial correlation in the spin angular momentum distribution of galaxies by analyzing the S/Z distribution of 9825 galaxies to constrain the galaxy formation scenarios but found no significant trend.  All these studies were performed by visual inspection of image data.

We are exploiting such studies by compiling a much larger database for S/Z spin discrimination using recent collection of high-resolution images of galaxies, such as the SDSS, PanSTARRS, and Subaru Hyper Suprime-Cam Public Data Release 2. To make robust catalogs, we are developing deep learning software that will allow unbiased, automatic S/Z classification, which will be reported in a forth-coming paper.  The current paper forms the first basis of our series of studies using S/Z parity to study the distribution and origin of galaxy spin vector.

Several papers report on the distribution of the position angles of the major axes of disk galaxies and/or the axis ratio distribution of galaxies to check for departures from a random distribution (e.g., \citet{Flin1990}, \citet{Navarro2004}, \citet{Trujillo2006}, Lee et al. 2007, Bett et al. 2007). These approaches involve quantitative measurements of the position angle and/or axis ratio, which are not free from measurement errors and could be subject to systematic error owing to the method employed. 

The present method uses only three pieces of information (S or Z, which side of the minor axis is nearer to us, and which side of the major axis is approaching) and does not involve quantitative errors in measurement.  Therefore, we think that it provides a more robust comparison with theoretical models than using the position angle distribution or the axis ratio distribution.  We emphasize that our approach using the spiral winding direction as described in this paper is independent of and complementary to approaches (e.g., \citet{Flin1990}, \citet{Navarro2004}, \citet{Trujillo2006}, \citet{Lee2007}, \citet{Bett2007} using the position angle and/or axis ratio distributions to study the spin vorticity distribution of galaxies in the Universe.

\citet{Hayes2017}, \citet{Shamir2017a}, and \cite{Shamir2017b} argued for the S/Z asymmetry observed in the Galaxy Zoo survey and SDSS. Our second forthcoming paper will report on the technique and early results of S/Z classification by deep learning of about 90,000 HSC sample galaxies observed by the Subaru Telescope.

Observational mapping of the microwave background and studying the large-scale distribution of galaxies have been very successful in constraining the expansion model of the Universe.  These two complementary approaches provided a glimpse of the evolution of the scalar density distribution of the Universe.  They were strong tools for understanding the formation of the structure of the Universe.  The present paper provides a new basis for examining the distribution of galaxy spins in the Universe using only an imaging study without recourse to spectroscopy.  Mapping the vector fields, velocity field, and vorticity vector field might provide new insights into the history of the Universe.

\section{Conclusion}

We compiled a spin catalog of 146 spiral galaxies for which the sense of winding of the spiral structure (S/Z), the dust-lane-dominant side of the minor axis, and the approaching side of the major axis are observationally confirmed.  The catalog shows a perfect correlation among these three pieces of information suggesting strongly that all the spirals are trailing provided that the dust-lane-dominant side of disk galaxies is closer to us which is generally assumed but has been demonstrated by means of differential reddening of globular cluster only for M31 (\citet{Iye1985}).

The present study provides a basis for using the S/Z winding direction of spiral galaxies to study the spin parity distribution of galaxies in order to constrain the formation scenarios of galaxies and the structure of the Universe.


\acknowledgments
 
The authors are grateful to the anonymous referee for his/her careful comments on the original version of this paper. The useful comments of Dr. Masafumi Yagi on an early manuscript are also acknowledged.

\end{document}